# Coexistence of Multiple Distinct Corner-State Classes via Nearest-Neighbor Couplings Alone

Luhong Su[1,4†], Dongyi Wang[3†*], Shenglong Guo[2], Xinjiang Chen[2], Lei Zhou[4,5*], Shaojie Ma[1,4*], and Mengyao Li[2*]

[1] Shanghai Engineering Research Centre of Ultra Precision Optical Manufacturing, College of Future Information Technology, Fudan University, Shanghai 200433, China

[2] Institute of Materials Research, Tsinghua Shenzhen International Graduate School, Tsinghua University, Shenzhen 518055, China

[3] New Cornerstone Science Laboratory, Department of Physics, The University of Hong Kong, Hong Kong 999077, China

[4] Shanghai Key Laboratory of Metasurfaces for Light Manipulation, Shanghai 200433, China

[5] State Key Laboratory of Surface Physics and Key Laboratory of Micro and Nano Photonic Structures (Ministry of Education) and Department of Physics, Fudan University, Shanghai 200438, China

[*]Corresponding Authors: Dongyi Wang, Email: physwang@hku.hk

Lei Zhou, E-mail: phzhou@fudan.edu.cn

Shaojie Ma, Email: shaojiema@fudan.edu.cn

Mengyao Li, E-mail: mengyaoli@sz.tsinghua.edu.cn

†These authors contributed equally to this work

**The coexistence of multiple distinct localized-state classes, including different types of topological corner states, is highly desired; yet its realization typically requires sufficiently strong long-range couplings, bringing challenges to experimental realizations across a wide range of systems. Here, we demonstrate the emergence of multiple localized corner modes evolution under purely nearest-neighbor coupling conditions without long-range couplings, not even next-nearest-neighbour couplings, and experimentally confirm their coexistence in a single lattice. Our platform simultaneously hosts type-I and type-II corner states, corner localized bound states in the continuum, flat-band states, and edge states. We experimentally realize such lattice using topolectrical circuits, a system that can strictly exclude long-range coupling by design, enabling direct characterizations of both the energy-band spectrum and the spatial profiles of the localized eigenmode. This work demonstrates that solely modifying nearest-neighbor couplings can replace the role**

**of long-range interactions in realizing multiple types of corner states, providing a deterministic and widely accessible route to multiple corner-state classes and rich localized states, with immediate potential for mode-division multiplexing and multimode topological devices.**

Recent advances in higher-order topological band theory have fundamentally transformed the landscape of bulk-boundary correspondence, leading to the discovery of topological corner states. In a two-dimensional (2D) system, such topological states manifest as bound states located at the zero-dimensional boundary of higher-order topological phases (HOTPs)[1–14].This unique localization has inspired several novel applications including quantum optics[15], thermal/acoustic engineering[16–18], and topological lasers[19,20]. Numerous works have been devoted to exploring 2D second-order topological insulators, which exhibit distinct classes of topological corner states through judiciously designed coupling schemes. Among these, type-I corner states located exactly at the geometric corner are the most commonly observed variety [21–23]. In parallel, type-II corner states, which reside in the units immediately adjacent to the exact corner, have also been demonstrated[13]. Beyond corner states in HOTPs, other classes of localized states have also attracted considerable attention. Bound states in the continuum (BICs) enable a crucial spatial confinement that traps energy in the continuum without radiating through symmetry protection or destructive interference[24–29], while flat-band compact localized state (CLS) achieves perfect finite-range confinement as a consequence of destructive interference in dispersionless bands. These two paradigms provide complementary mechanisms for wave localization beyond conventional topological confinement.

The coexistence of multiple classes of localized states is of particular interest, as it offers promising pathways for mode-division multiplexing, multi-channel signal processing, and richer topological functionalities. However, almost all existing proposals that generate multiple types of corner states rely on the introduction of sufficiently strong long-range couplings[13,14,30–32], like next-nearest-neighbour (NNN) couplings[9]. Such coupling configurations make experimental realization challenging across a wide range of natural systems where long-range couplings are limited by natural exponential decay of electronic wave functions, and their precise tuning and controlling remains demanding. Concurrently, the existence of such couplings compromises the exact flatness upon which CLSs rely, imposing additional constraints on the practical exploitation of flat bands. Given these inherent difficulties, the requirement on long-range coupling strengths and increased complexity in lattice design remain a significant obstacle to extending these phenomena to practical applications[14,16–18]. Thus, achieving a realistic single-lattice platform entirely free of long-range couplings that can simultaneously host multiple classes of topological corner states and other localized states like BICs and flat-band CLSs, is of practical desire.

In this letter, we demonstrate that a lattice with purely nearest-neighbor couplings, can simultaneously host these multiple distinct classes of localized states. Unlike previous approaches, our mechanism does not require any long-range couplings in the lattice. Instead, the effective mode coupling behavior emerges naturally through second-order scattering processes in the mode space. Our approach leverages the engineered coupling condition by trimerizing the Kagome lattice unit cells, forming a larger unit cell where the intercell, trimerization strength, and breathing Kagome intracell couplings are tuned over a broad range to access different topological phases, the associated band structure, and bound state properties. This configuration gives rise to three distinct types of corner bound states within the same lattice: type-I corner states localized at the outermost corners, even- and odd-symmetric type-II corner states confined at the units adjacent to the exact corner, and corner BICs emerging under certain coupling-ratio conditions. Meanwhile, the Kagome flat band is also preserved, hosting CLSs. Using topolectrical circuits [33–40] which by design are free of long-range couplings, we directly measure both the impedance spectrum and the spatial profiles of these localized eigenmodes, providing unambiguous experimental evidences for the coexistence of multiple distinct corner-state classes. Our work demonstrates that the tunable degrees of freedom offered solely by a suitably engineered nearest-neighbor couplings lattice can effectively replace long-range interactions in simpler lattice configuration, provides a deterministic route to realize the coexistence of multiple localized states including different types of topological corner states. This finding offers a simple and experimentally accessible route toward multifunctional topological lattices and opens new possibilities for mode-division multiplexing in topological photonics and phononics.

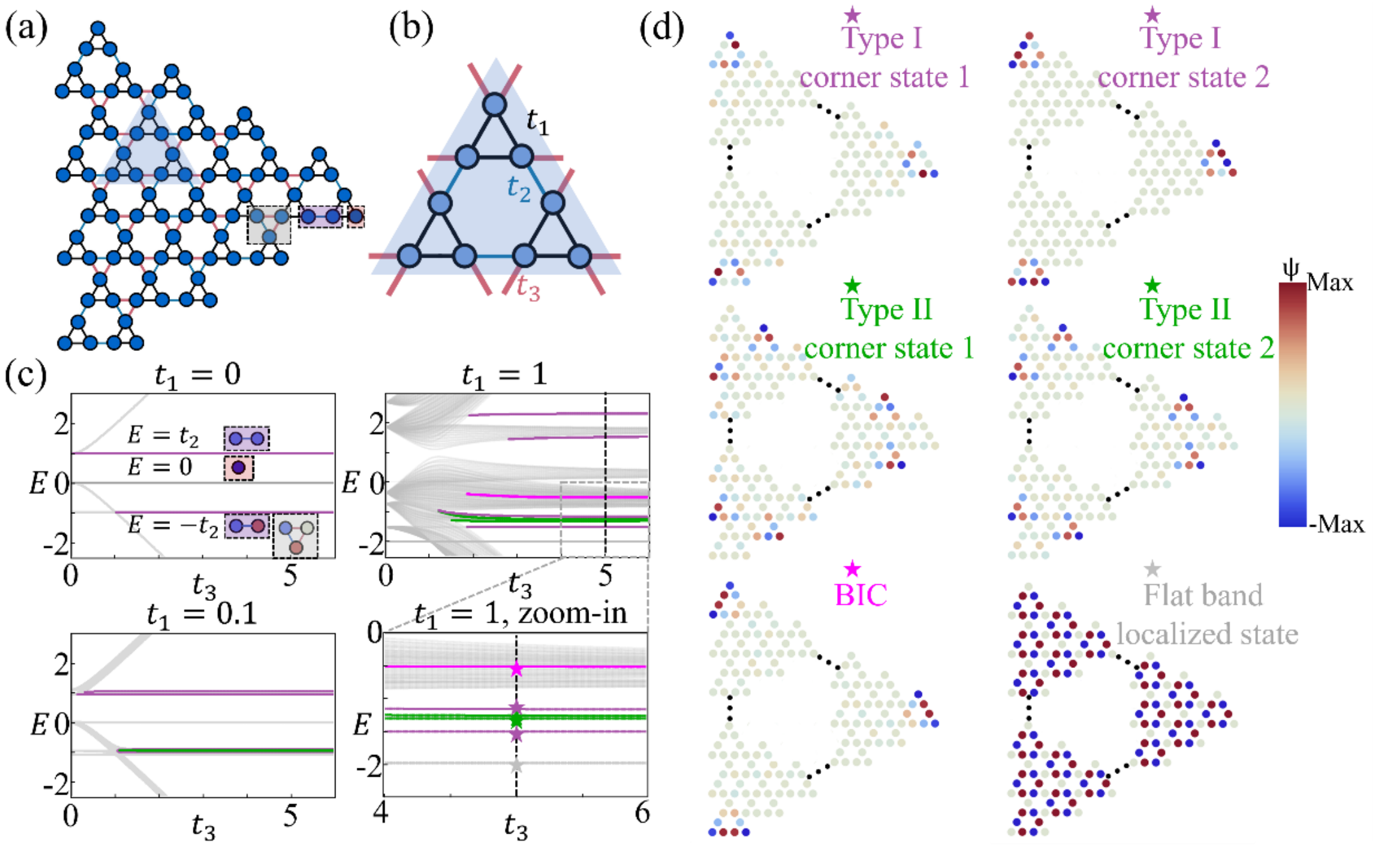

**FIG. 1. The lattice design and the supported multiple localized states**. (a) Schematic of the trimerized Kagome lattice with three sets of coupling: intra-cell couplings $t_1$ forming bonds within each small triangle (black line), intra-cell couplings $t_2$ connecting neighboring small triangles inside the same unit cell (blue line), and inter-cell couplings $t_3$ connecting adjacent unit cells (pink line). Each unit cell contains three small triangles, comprising nine sites in total. (b) The zoomed-in view of a nine-site unit cell. (c) Calculated band structures for the lattice shown in (a) containing 12 unit-cells along each crystal axis, with fixed $t_2 = 1$, $t_3$ ranging from 0 to 6 and different $t_1$ values ($t_1$= 0, 0.1, 1, as indicated). The zoomed-in panel highlights the eigenvalues associated with localized eigenstates, with purple, green, magenta, and gray stars denoting type-I corner states, type-II corner states, BICs and flat-band CLSs, respectively. (d) Theoretical spatial profiles of representative localized modes for $t_1 = 1$, showing distinct localization characteristics of these states.

The system under investigation is a variation of the trimerized Kagome lattice, as shown in Fig. 1 (a), whose unit cell consists of three coupled triangles containing nine sites in total. The lattice is trimerized through three distinct couplings: intra-cell couplings $t_1$ bonds sites within each small triangle, intra-cell couplings $t_2$ connects different small triangles within the same unit cell, and inter-cell couplings $t_3$ links adjacent unit cells [Fig. 1 (b)]. The system can be described by the tight-binding model (TBM) Hamiltonian:

$$H = \sum_{<i,j>} t_{ij}\, c_i^{\dagger} c_j + \text{h.c.} \tag{1}$$

where $t_{ij} = t_1$ or $t_2$ for intra-cell bonds and $t_{ij} = t_3$ for inter-cell bonds. In this geometry, the localization of eigenstates can be flexibly tuned through the interplay among these three coupling strengths.

We investigate the evolution of the band structure against the trimerization captured by intracell couplings $t_1$ and inter-cell couplings $t_3$ while fixing intra-cell couplings $t_2 = 1$, with the results presented in Fig. 1 (c). Starting from the isolated limit $t_1 = 0$, the lattice exhibits a large number of flat bands with vanishing group velocity pinned at $E = 0$ and $E = \pm t_2$. Each of these flat bands host degenerated CLSs, including isolated single-site monomers at $E = 0$, two-site dimers at $E = t_2$ (symmetric) and $E = -t_2$ (antisymmetric), and three-site coupled states at $E = -t_2$ consisting of an antisymmetric dimer on two sites connected by $t_2$ with the third site having zero field amplitude, as shown in the insets. When a finite coupling $t_1$ is introduced, these localized modes hybridize. Consequently, most of the initially localized states can no longer sustain stable standing-wave patterns within a confined area and thus evolve into extended and dispersive bulk states. This delocalization is accompanied by a progressive broadening of their eigenfrequencies with increasing of $t_1$, which develop finite bandwidths, a trend accurately captured by the gray curves in Fig. 1 (c). Remarkably, once $t_3$ exceeds a sufficiently large threshold, a subset of modes exhibits eigenfrequencies that become asymptotically insensitive to $t_3$, indicating their localization within individual unit cells. As summarized in Fig. 1 (d), these $t_3$-robust eigenfrequencies correspond to several distinct types of localized states, including two sets of type-I corner states localized at the terminal corner, two sets of type-II corner states localized at the inner corner being even-symmetric and odd-symmetric respectively, one set of BICs and one set of flat-band CLSs. These modes are respectively marked by purple, green, magenta, and gray stars in the zoomed-in lower panel of Fig. 1 (c), which we detailly discuss in the following paragraphs.

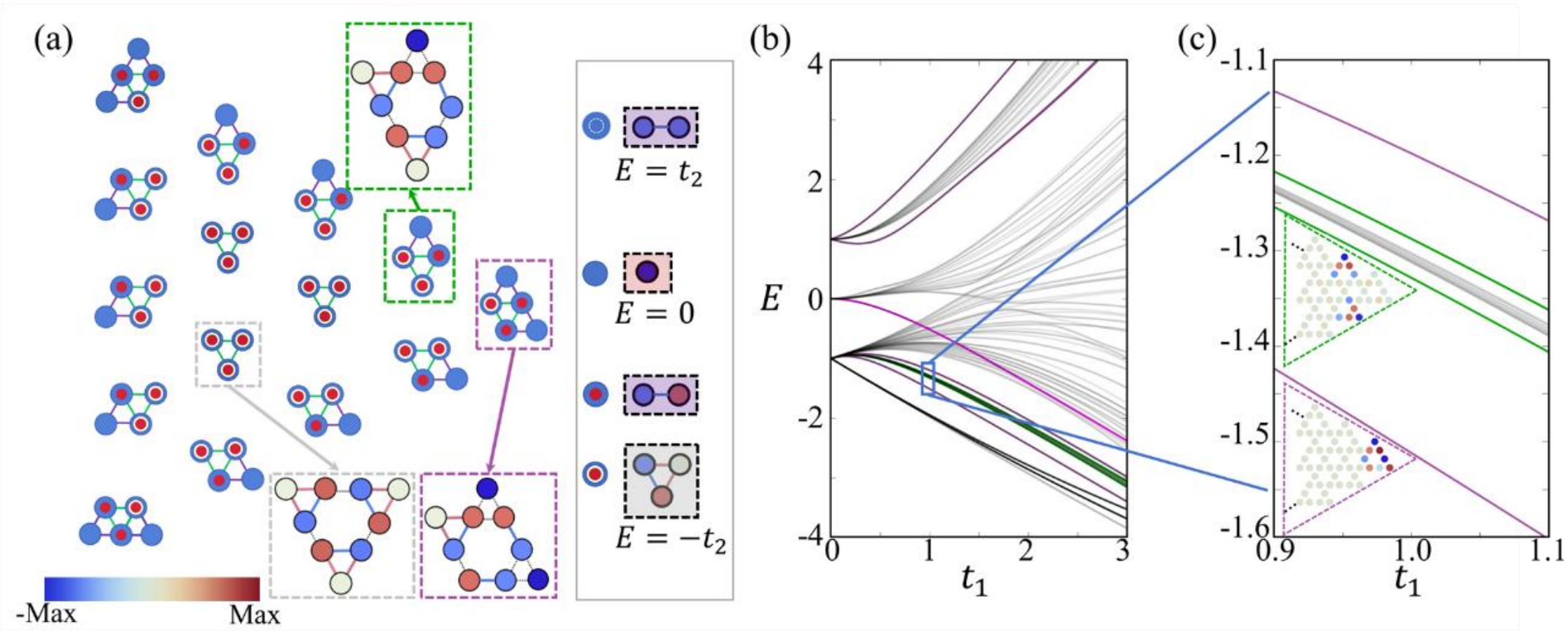

**FIG. 2. Mode analysis in Hilbert eigenmode subspace for the formation of multiple localized states**. (a) Projection of the coupled trimerized Kagome lattice at $t_1 \neq 0$ onto the Hilbert eigenmode subspace spanned by the four parent CLSs at $t_1 = 0$. The legends illustrate the correspondence between the real-space parent CLSs (monomer, dimer and three-site coupled states for $E = 0, \pm t_2$ at the $t_1 = 0$ limit) and their mode-space representations: colors denote real-space distributions and symbols assign each mode a unique label within the mode-space. The insets, enclosed by dashed purple, green, and gray boxes, display the real-space counterparts of the individual, mutually decoupled mode-space substructures, spatially located at the corners, boundaries, and bulk interior, respectively. Modes at $E = t_2$ omitted for clarity, as they overlap spatially with $E = -t_2$ modes on the boundary. (b) Band structures reconstructed analytically from spectra in the Hilbert eigenmode subspace via second-order perturbation theory, fixing $t_2 = 1$, $t_3 = 5$, and varying $t_1$ from 0 to 3. Purple, green, and magenta curves correspond to spectra of type-I corner states, type-II corner states, BIC modes, respectively. (c) Zoomed-in view of the band structures showing two sets of type-I and type-II corner states.

We further analyze these modes individually in the mode space. Considering that $t_1$ and $t_2$ are comparable in magnitude while much smaller than $t_3$ to be treated as perturbation (i.e., $t_3 \gg t_1, t_2$), the low-frequency eigenstates can be effectively described by the coupling of four-parent CLSs at the trimerized $t_1 = 0$ limit, as shown in Fig. 2 (a). Although the trimers connected by $t_2$ and $t_3$ also contain high-energy modes around the order of $t_3$, their influence becomes negligible in this strongly trimerized regime.

Thus, we first track the evolution of the states at energy $E = -t_2$ in the $t_1 = 0$ isolated limit. These states originate from two distinct cluster mode: a two-site dimer located at the boundary and a three-site coupled state residing in the bulk. In both cases, the eigenmode is localized on the two sites connected by $t_2$, with equal field amplitudes and opposite phases. Projecting the full lattice Hamiltonian onto the corresponding cluster Hilbert eigenmode subspace yields the coupling network shown by the isolated small triangles in Fig. 2 (a), green and purple lines denoting different effective couplings in the Hilbert subspace. Notably, in the three-site configuration, destructive interference within each unit cell, a hallmark of the corner-sharing triangular geometry inherent to the Kagome lattice, forces a strictly zero field on the $t_3$-connected site, causing the effective coupling between neighboring unit cells to vanish constantly, even when $t_1 \neq 0$. As a result, these modes remain perfectly localized and form the conventional Kagome flat band at $E = -(t_1 + t_2)$, whose degeneracy equals the number of unit cells.

Furthermore, at the boundaries and corners, the fully isolated structure at $t_1 = 0$ supports two additional types of localized modes: a single-site monomer mode with energy $E = 0$ and a symmetric in-phase dimer mode with energy $E = +t_2$. When $t_1 \neq 0$, these modes remain effectively decoupled between different unit cells. Nevertheless, inside each unit cell, these additional modes participate in the hybridization, allowing the system in Hilbert space to exhibit distinct architecture, especially in the edge and corner regions. To be specific, three different kinds of localized modes are formed as shown in Fig. 2 (a), being corner localized modes (denoted by dashed purple boxes), edge localized modes (denoted by dashed green boxes) and bulk localized modes (denoted by dashed gray boxes), all connected by nonzero $t_1$.

Therefore, the entire low-energy spectrum can be reconstructed by analyzing the eigenvalues of these localized coupling structures, as shown in Fig. 2 (b). This approach accurately predicts the energies and spatial profiles of the type-I corner-localized modes and boundary-localized modes (see more details in Supplementary Information Section A). It's also noted that among the type-I corner-localized modes, a special subset resides in the continuous bulk of the spectrum rather than in a gap. Nevertheless, it maintains spatially localized at the corner, as demonstrated by the magenta curve, thus identified as a BIC. The validity of this physical picture is confirmed by the excellent agreement between the analytically reconstructed spectra and the full numerical calculations.

Notably, high-energy modes can mediate an effective coupling among low-energy modes via a second-order process: a low-energy mode (with characteristic energy scale $t_1, t_2$) couples to a high-energy mode near $\pm t_3$ through the hopping $t_1$, and subsequently scatters back into another low-energy mode through a second hopping event. This scattering mechanism generates an effective long-range-coupling network within the low-energy mode space, with coupling strength on the order of $t_1^2/t_3$, despite the fact that only nearest-neighbor couplings exist in the original real-space lattice. As a result, two branches of the boundary states experience a slight energy shift near the corner due to its coupling to corner localized states, giving rise to a new class of states localized in the vicinity of the corner. These states are commonly referred to as the type-II corner states, which are further distinguished by even and odd symmetry, as explicitly demonstrated by the green lines in Fig. 2 (c). Unlike conventional approaches that rely on explicit long-range couplings in real space, our approach produces multiple corner states using only nearest-neighbor couplings. The effective long-range behavior emerges naturally through scattering processes in Hilbert space, providing a significantly more experimentally accessible route toward realizing different types of coexisting higher-order localized states.

To experimentally realize and characterize these multiple types of localized topological corner states, we employ a versatile platform which maps the TBM Hamiltonian directly onto measurable circuit Laplacian (see Supplementary Information Section B). Topolectrical circuits allow precise control of coupling ratios through capacitor values, and provide straightforward measurement of eigenmode profiles via voltage measurements. Each lattice site is mapped to a topolectrical circuit node, with the hopping amplitude $t_{ij}$ between neighboring sites $i$ and $j$ implemented by a capacitor $C_{ij}$ connecting the two nodes. Additionally, each node is grounded through an inductor $L$ and a capacitor $C_n$, together they set the reference frequency and correspondingly determine the on-site potential (see more details in methods).

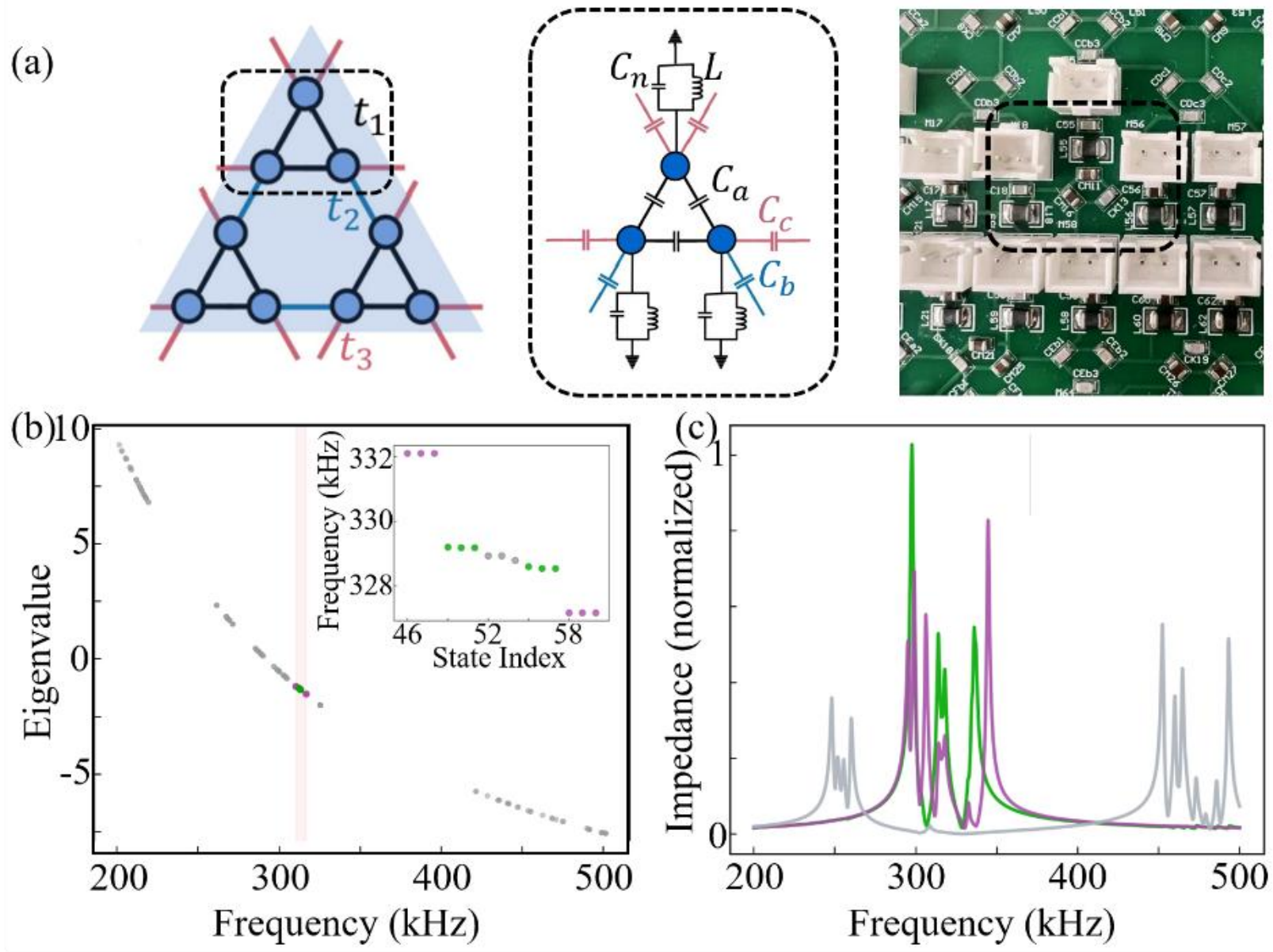

**FIG. 3. Topolectrical circuit implementation of the** trimerized Kagome lattice**.** (a) Schematics of the designed topolectrical circuits, with capacitors $C_a = C_b$ implementing equal intra-cell couplings $t_1 = t_2$, and capacitors $C_c$ realizing the inter-cell couplings $t_3$. Each node is grounded through an inductor $L$ and a capacitor $C_n$. (b) Eigenvalues and corresponding resonant frequencies of the topolectrical circuits. The inset displays a magnified view of the in-gap region containing Type-I (purple dots) and Type-II (green dots) corner states. (c) Numerically calculated impedance spectra of the lossless topolectrical circuit consisting of 135 sites. The purple, green and grey curves denote the impedance responses of sites hosting Type-I corner states, Type-II corner states and bulk states, respectively.

We implement the trimerized Kagome model using a topolectrical circuit, as illustrated in Fig. 3 (a). A finite triangular lattice containing 5 unit-cells along each crystal axis (135 sites in total) is designed. The intra-cell coupling $t_1 = t_2$ of the trimerized lattice is implemented by capacitance $C_a = C_b = 2$ nF, while the inter-cell coupling $t_3$ is set by $C_c = 10$ nF, yielding a ratio of $t_1 : t_3 = 1:5$ that places the system deep in the phase region hosting multiple types of localized states [Fig. 1 (c)]. Each node is grounded through an inductor $L = 10\mu$H and a capacitor $C_n = (26 - m_a \cdot C_a - m_b \cdot C_b - m_c \cdot C_c)$ nF, where $m_a, m_b$ and $m_c$ denote the number of intra- and inter-cell couplings connected to node $n$ to ensure a uniform total capacitor for each node of $C_{tot} = 26$ nF. The resonance frequency is set at $f_0 = 1/(2\pi\sqrt{LC_{tot}}) \approx 300$ kHz. Fig. 3 (b) presents the eigenvalues and corresponding resonant frequencies of the trimerized Kagome lattice circuit, where the red-shaded region highlights the frequency domain of interest containing Type-I and Type-II corner states, denoted by purple and green dots, respectively. Fig. 3 (c) shows the numerically calculated impedance spectra of the lossless topoletrical circuit with 135 sites. The purple, green and grey curves illustrate the impedance of Type-I corner states, Type-II corner states and bulk states, respectively.

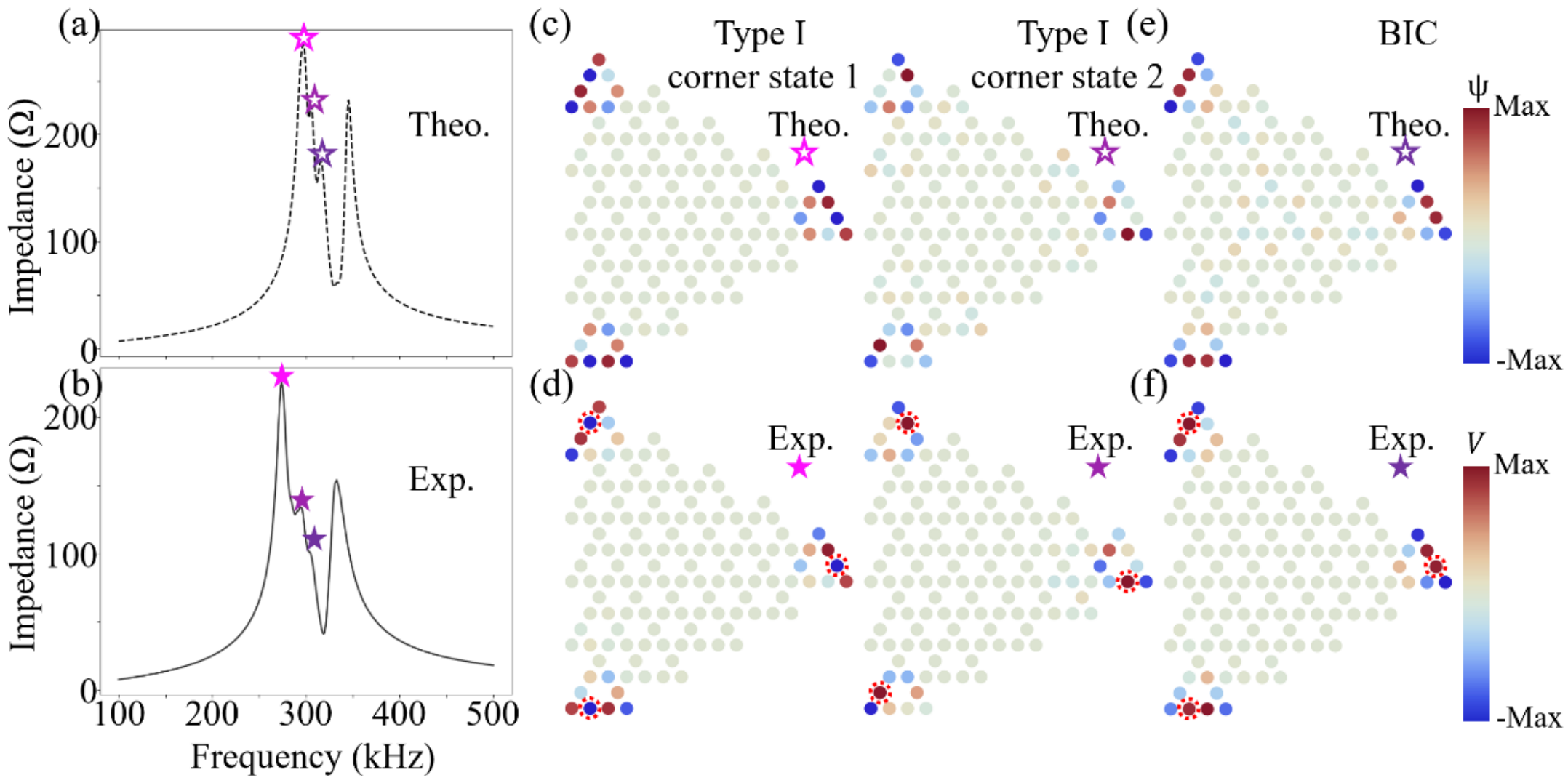

**FIG. 4. Theoretical and experimental characterizations of type-I corner states, including a set of BIC modes.** (a, b) Theoretically predicted (a) and experimentally measured (b) impedance $Z_{aa}$ spectra based on lossy circuit Laplacian, with excitations at exact corner unit cells. Resonant peaks at distinct frequencies reveal different sets of type-I corner states (denoted by stars with different colors), including a set of BIC corner modes. (c-f) Theoretically predicted (c, e) and experimentally measured (d, f) voltage distributions for three representative type-I corner states, each localized at the exact corners of the triangular lattice, yet possessing different spatial profiles of amplitudes and phases. Panels (e, f) show the BIC corner mode. Hollow and solid stars denote theoretically predicted and experimentally observed localized states, and dashed red circles mark the excitation position (see Supplementary Information Sections C and D for more experimental details).

The impedance spectrum is measured using an impedance analyzer of $Z_{aa}$, with a single probe placed at various corner positions. The corresponding results, shown in Fig. 4 (b) are well-consistent with theoretical predictions [Fig. 4 (a)]. Distinct resonant peaks appear at different frequencies, each corresponding to different corner state sets. Two resonances at 295 kHz and 315kHz located within the bulk band gap are identified as type-I corner states, when the probe is placed at the exact corner unit cell. Additionally, a high-quality resonance at 275 kHz is observed inside the bulk band, corresponding to type-I corner BICs. It should be noted that, owing to parasitic resistance, the experimental linewidths are significantly broader than the ideal ones in Fig. 3 (c). However, including parasitic resistance in the theoretical calculations yields linewidths that match the experimental results, reinforcing the consistency between theory and experiment.

To definitively characterize the exact type of the measured corner states, spatial voltage distributions across the lattice are mapped for each resonant mode. Following standard techniques for topolectrical circuits: we excite the topolectrical circuit at the eigenmode frequency and measure the voltage amplitude at every node using a lock-in amplifier. Fig. 4 (d) illustrate the measured spatial profiles for two in-gap type I corner modes at different frequencies. Both of them exhibit maximum amplitude at the outermost corner unit cell, yet possessing distinct amplitude and phase distribution profiles that closely match with the theoretical predictions [Fig. 4 (c)], confirming successful generation of conventional corner localization. Fig. 4 (f) presents the measured profile of the BIC corner mode, also consisting well with theoretical predictions [Fig. 4 (e)]. Despite its energy lying within the bulk bands, such BIC modes exhibit strong corner localization with quality factors exceeding 30, which is over ten times higher than the peaks of other type I corner modes. Moreover, the fourth peak in Fig. 4 (a, b) corresponds to the flat band. While these modes are also localized at the corners, they are not topological corner ones (see more details in Supplementary Information Section E).

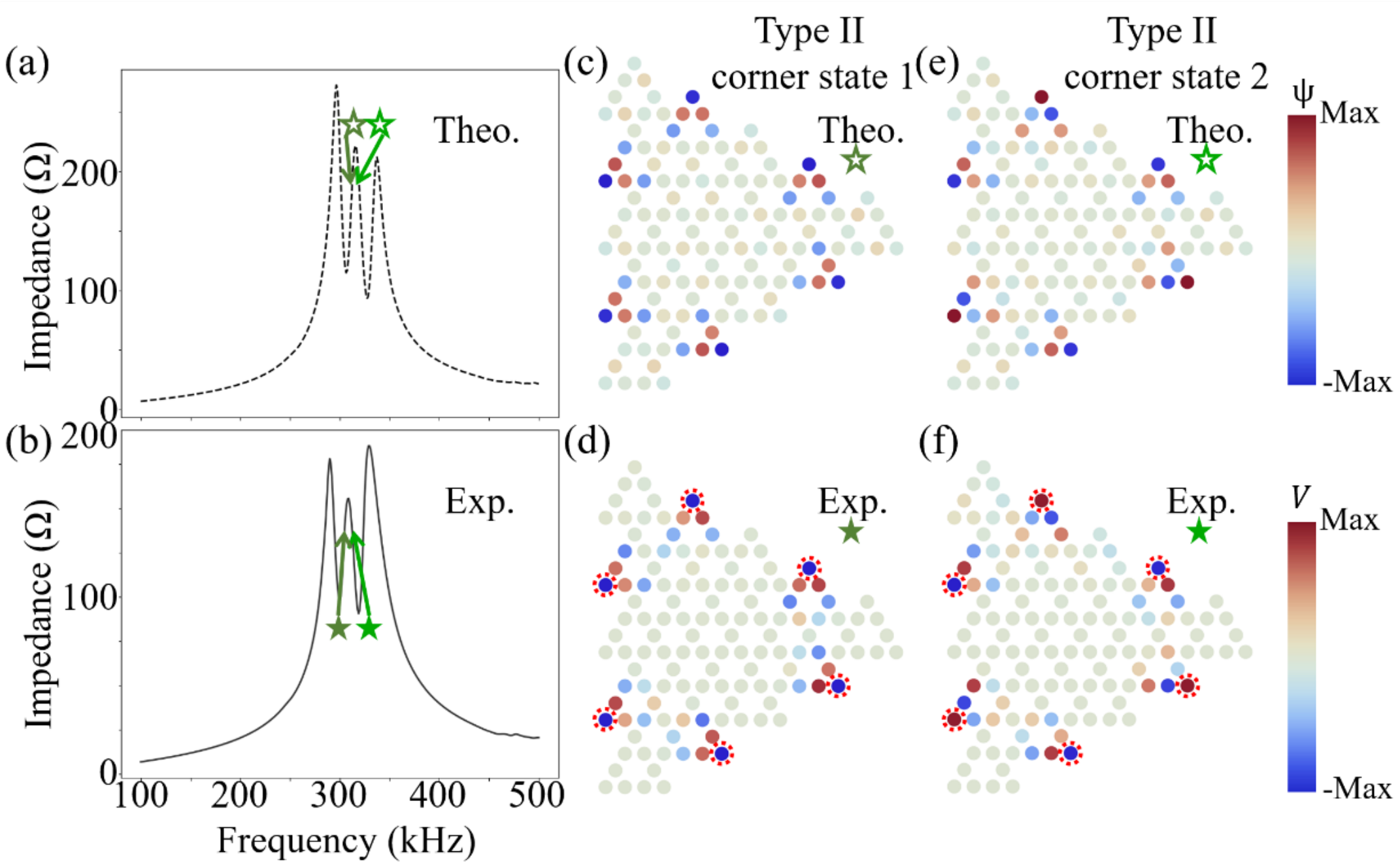

**FIG. 5. Theoretical and experimental characterizations of type II corner states.** (a, b) Theoretically predicted (a) and experimentally measured (b) impedance $Z_{aa}$ with excitation at unit cells next to exact corners revealing resonant peaks at distinct frequencies for type-II corner states. (c-f) Theoretically predicted (c, e) and experimentally measured (d, f) voltage profiles for (c, d) even-symmetric and (e, f) odd-symmetric type-II corner states, exhibiting amplitudes confined within unit cells next to the exact corner. Hollow and solid stars denote localized states captured theoretically and experimentally, dashed red circles denote excitation position.

Fig. 5 displays the experimental characterizations of in-gap type-II corner states, featuring the second peak in the spectra [Fig. 5 (a, b)]. As shown in Fig. 5 (b), when the excitations are applied at the unit cells next to the exact corner, clear resonant peaks of in-gap type II corner state appear between the type-I corner state peaks, in excellent agreement with theoretical predictions [Fig. 5 (a)]. The key features of these states: strong confined localization at the unit cells next to the exact corner, with neighbor units exhibiting in-phase (even symmetric) or out-of-phase (odd-symmetric) patterns for two types of corner states, are observed experimentally [Fig. 5 (d, f)], consisting well with theoretical predictions [Fig. 5 (c, e)]. Moreover, the first peak in Fig. 5 (a, b) corresponds to the bulk band exhibiting edge states, while the third peak belongs to the flat band exhibiting also modes localized at unit cells next to the exact corner but not recognized as topological corner ones (see more details in Supplementary Information section E).

## Discussions and Conclusions

In conclusion, we have experimentally demonstrated multiple classes of localized states in a trimerized Kagome lattice by topolectrical circuits. Our platform supports type-I corner states, even- and odd-symmetric type-II corner states, flat-band CLSs, edge states, and corner BICs — all coexisting within the same band structure without requiring long-range couplings, such as NNN couplings or beyond. Direct measurements of both the admittance spectrum and spatial mode profiles confirms the distinct nature of each corner state class. Our work establishes a versatile platform for exploring multi-mode topological phenomena and opens new avenues for mode-division multiplexing in topological photonic and phononic devices. Critically, the engineered coupling condition inherent to the trimerized geometry provides the mechanism for this multiplicity, as revealed by projection onto the Hilbert eigenmode subspace, offering an alternative to conventional coupling-inversion schemes that rely on stringent long-range coupling conditions. Furthermore, this Hilbert-subspace framework provides a route to more intricate lattice geometries, unlocking new tunable degrees of freedom for mode control under purely nearest-neighbor coupling conditions. The coexistence of multiple corner-state classes in a single lattice may benefit applications requiring spatially multiplexed energy confinement, such as multimode corner-emitting lasers. More broadly, our results highlight the untapped potential of the engineered coupling condition for generating rich topological phase diagrams in experimentally accessible systems.

## REFRENCES

## ACKNOWLEDGMENTS

We thank financial support from the National Key R&D Program of China (Grant No. 2023YFA1407700, 2023YFA1406901, 2023YFB3811400), National Natural Science Foundation of China (Grant No. 12374343, 12404435), Guangdong Science and Technology Commission (Grant No.2025A1515010774), and the start-up funding of Fudan University (JIH1232133Y)

**Authors' contributions**

Luhong Su and Dongyi Wang contributed equally to this work. M.L. and S.M. conceived the idea. L.S., D.W. and S.G. performed theoretical derivations. L.S. designed the circuits systems. L.S., D.W., S.G., and X.C. carried out experimental testing. L.S. and D.W. composed the original draft. S.M., L.Z. and M.L. revised the draft. S.M., L.Z., D.W. and M.L. supervised the project. All authors contributed to the discussion and preparation of the manuscript.

**Competing interests**

The authors declare no conflict of interest.

**Additional information**

**Data and materials availability:** The data that support the findings of this study are available from the corresponding author upon reasonable request.

**Methods**

In topolectrical circuit design, the tight-binding model is mapped to a topolectrical circuit captured by the circuit Laplacian [see details in Supplementary Information section B]. According to Kirchhoff's law,

$$I_i = J_{ii}V_i + \sum_{i\neq j} J_{ij}V_j \tag{2}$$

$I_i$ donates the current at node $i$ and $V_{i(j)}$ donate the voltages of node $i(j)$, $J_{ii}$ is the sum of admittances induced by all capacitors/inductors connected to node $i$, and $J_{ij}$ (with i≠j) is the negative admittance induced by capacitor for coupling between nodes $i$ and $j$. Due to the inherit features of capacitors and inductors, Eq. (2) can be rewritten as

$$I_i = i\omega(C_n + \sum_{i\neq j} C_{ij} - \frac{1}{\omega^2 \mathrm{L}})V_i + \sum_{i\neq j} i\omega C_{ij}V_j \tag{3}$$

$\sum_{i\neq j} C_{ij}$ denotes the sum of all the capacitors connected to node $i$ for lattice couplings. By suitably choosing the grounding capacitor $C_n$, the on-site energies are tuned to be identical across all nodes. The TBM Hamiltonian is directly mapped to the circuit Laplacian $J$. Consequently, the admittance eigenvalues $j_n$ of $J(\omega)$ correspond directly to the eigen energies of the TBM Hamiltonian, with admittance modes corresponding to topological states possessing certain energies $j_n$.